\documentclass[preprint]{elsarticle}
\usepackage{graphicx}
\usepackage{dcolumn}
\usepackage{amsmath}
\usepackage{amssymb}
\usepackage{bm}

\begin{document}

\def\vri{\vec{r}_{i}}
\def\vrj{\vec{r}_{j}}
\def\rij{r_{ij}}
\def\vrij{\vec{r}_{ij}}
\def\drij{\hat{r}_{ij}}
\def\vdr{\delta\vec{r}}
\def\dr{\delta{r}}
\def\s{\hat{s}}

\title{Distribution of volumes and coordination number in
  jammed matter: mesoscopic ensemble}



\author[focal]{Ping Wang}
\author[focal]{Chaoming Song}
\author[focal]{Yuliang Jin}
\author[focal]{Kun Wang}
\author[focal]{Hern\'an A. Makse\corref{cor1}}
\ead{hmakse@lev.ccny.cuny.edu}
\cortext[cor1]{Corresponding author}
\address{Levich Institute and Physics Department, City College of New
  York, New York, NY 10031, US}

\date{\today }

\begin{abstract}
    We investigate the distribution of the volume
    and coordination number associated to each particle in a jammed
    packing of monodisperse hard sphere using the mesoscopic ensemble
    developed in Nature {\bf 453}, 606 (2008). Theory predicts an
    exponential distribution of the orientational volumes for random
    close packings and random loose packings. A comparison with
    computer generated packings reveals deviations from the
    theoretical prediction in the volume distribution, which can be
    better modeled by a compressed exponential function. On the other
    hand, the average of the volumes is well reproduced by the theory
    leading to good predictions of the limiting densities of RCP and
    RLP. We discuss a more exact theory to capture the volume
    distribution in its entire range. The available data suggests a
    plausible order/disorder transition defining random close
    packings. Finally, we consider an extended ensemble to calculate
    the coordination number distribution which is shown to be of an
    exponential and inverse exponential form for coordinations larger
    and smaller than the average, respectively, in reasonable
    agreement with the simulated data.
\end{abstract}

\maketitle

\section{Introduction}

Jammed matter refers to a broad class of physical many-body
systems ranging from granular matter to frictionless emulsions,
and colloids. These systems share the property that their
constitutive particles can be blocked in a configuration far from
thermal equilibrium when undergoing a jamming transition.  The
statistical mechanical description of these materials is based on
the volume fluctuations of the system \cite{sirsam} taken to be
the conservative quantity instead of energy, as typically done in
thermal system. Therefore, the probability distribution of the
volume occupied by each jammed particle is of particular interest
and many studies have been devoted to investigate them in
detail \cite{Finney1970,aste,dauchot,jasna,Frenkel2008,Schroter2005,Brujic2007}.

Recently, a theory of volume fluctuations has been developed at
the mesoscopic level providing a relation between the volume
occupied by each particle and its number of contacts
\cite{jamming2}. Thus, the distribution of particle volumes in the
system is intimately related to the distribution of contacts per
particle. In this paper, we calculate the distribution of
(orientational) volumes occupied by each particle in a jammed
system of monodisperse hard spheres and the distribution of
coordination numbers by following the theoretical formalism of
\cite{jamming2}, which is in turn based on the Edwards statistical
mechanics of jamming \cite{sirsam}.

The distributions of volumes and contacts in real packings
represent ensemble averages in the statistical mechanics sense.
Therefore, the distributions depend on the state of the packings
specified by their compactivity through a Boltzmann-like
probability in the partition function. We show that in two
limiting cases of zero and infinite compactivity (corresponding to
the random close packing, RCP, and random loose packing, RLP,
respectively) the distributions can be obtained in analytical
form. Theory predicts that the distribution of orientational
volumes is exponential with a mean volume varying with the average
coordination number for RLP and constant for RCP.  The theoretical
predictions are compared with computer generated jammed packings
of equal-size spheres for any friction coefficient.  We find that
the theory well reproduces certain features of the numerical
distributions, but not all, in the entire range of volumes.

The mean value of the occupied volumes is well reproduced by the
theory. However, we find important deviations between theory and
simulations for the higher moments of the distribution.  For
instance, simulations show a plateau for small volumes while
theory predicts an exponential.  For intermediate values, the
exponential shape seems to provide a good fit to the simulated
data and the predicted dependence of the characteristic volume on
the coordination number follows partially the theoretical
prediction. However, a higher scrutiny shows deviations in the
tail of the distribution which is found to decay faster than
exponential, having a compressed exponential tail. We conclude
that the full understanding of the distribution requires a more
precise theory. We discuss how to obtain more exact solutions of
the volume distribution which can capture the behavior in the
entire ranges of volumes.

On the other hand, the distribution of coordination numbers provides fundamental information of the microscopic packing structures \cite{aste,Brujic2007, Clusel2009,Zhang2005,Troadec2002}, as well as important characteristics of the mesoscopic volume ensemble. In this paper, we study the distribution of coordination numbers by generalizing the ensemble proposed in \cite{jamming2} to include fluctuations in the number of contacts. Theory predicts an exponential decay for large coordination number and an inverse exponential for small coordination number.  Computer simulations well reproduce the predictions. Overall, the present paper serves as a critical assessment of the theoretical predictions of the mesoscopic theory towards the development of an exact formulation at the microscopic level that could capture the behavior in the entire range of volume and coordination number fluctuations of
jammed matter.

\section{Mesoscopic ensemble of jammed matter}
In this section, we briefly review the statistical mechanics theory developed at the mesoscopic level in \cite{jamming2}, which serves as the theoretical framework for the study of the distributions of volumes and coordination numbers. A theoretical formalism of the volume ensemble is the starting point for the statistical mechanics of jammed matter \cite{Edwards1989}. The role traditionally played by the energy in thermal systems is replaced by the volume, and a new parameter $X$, called ``compactivity", is introduced as an analogue of temperature. As a consequence, the canonical partition function can be written as:
\begin{equation}
\mathcal{Z}(X) = \int e^{-W/X} g(W)\Theta_W dW,
\label{eq:part1}
\end{equation}
where $W$ is the free volume function, $g(W)$ is the density of jammed states for a given volume $W$, and $\Theta_W$ imposes the jamming condition. It has been shown \cite{jamming2} that the free volume of coarse-grained ``quasiparticles" in a monodisperse hard sphere packing has an inverse relation with their coordination number $z$:
\begin{equation}
W(z) =  \frac{2\sqrt{3}}{z}V_g,
\label{eq:w1}
\end{equation}
where $V_g$ is the sphere volume. Since the quasiparticles are coarse-grained over a uniform background field produced by other particles, Eq.(\ref{eq:w1}) should be understood as a mean-field result at the mesoscopic level. Assuming the quasiparticles are independent, we can simplify the partition function Eq.(\ref{eq:part1}) by changing variables:
\begin{equation}
\mathcal{Z}_{\rm iso}(X) = \int_Z^6 e^{-W(z)/X} g(z)dz.
\label{eq:part2}
\end{equation}
The limit of integration here is given by the isostatic
condition \cite{alexander,moukarzel,edwards-grinev} over the
mechanical coordination number, $Z$ , counting the contacts
with nonzero forces. The mechanical coordination number is different from the geometrical
coordination number, $z$, which counts all contacts, even those with
zero forces. By definition, it is easy to see that the geometrical coordination number $z$ is always equal or greater than the mechanical coordination number $Z$, and in general we have $Z\le z\le 6$. The mechanical
coordination, $Z(\mu)$, depends on the friction, $\mu$, the interparticle friction coefficient, and varies between $Z(0)=2d=6$ and $Z(\infty)=d+1 = 4$ in dimensions $d = 3$. The density of states $g(z)$ is assumed to have an exponential
form, $g(z) = (h_z)^{z-2d} = e^{-(z-2d)/z^*}$, with
the constant $h_z\ll 1$, representing the typical separation of
the configurations in the phase space (analogous to the Planck
constant in quantum mechanics). We have $z^*=-1/\ln h_z$. Equation~(\ref{eq:part2}) provides a useful tool to calculate the ensemble average of any physical quantity $f(z)$ since
\begin{equation}
f(X,Z) = \frac{1}{\mathcal{Z}_{\rm iso}}\int_Z^6 f(z)e^{-W(z)/X} g(z)dz.
\label{eq:ensAve}
\end{equation}

\section{PDF of the orientational free volumes}

\begin{figure}
  \centering { \hspace{.6cm}\hbox {
      \resizebox{8cm}{!}{\includegraphics[angle=-90]{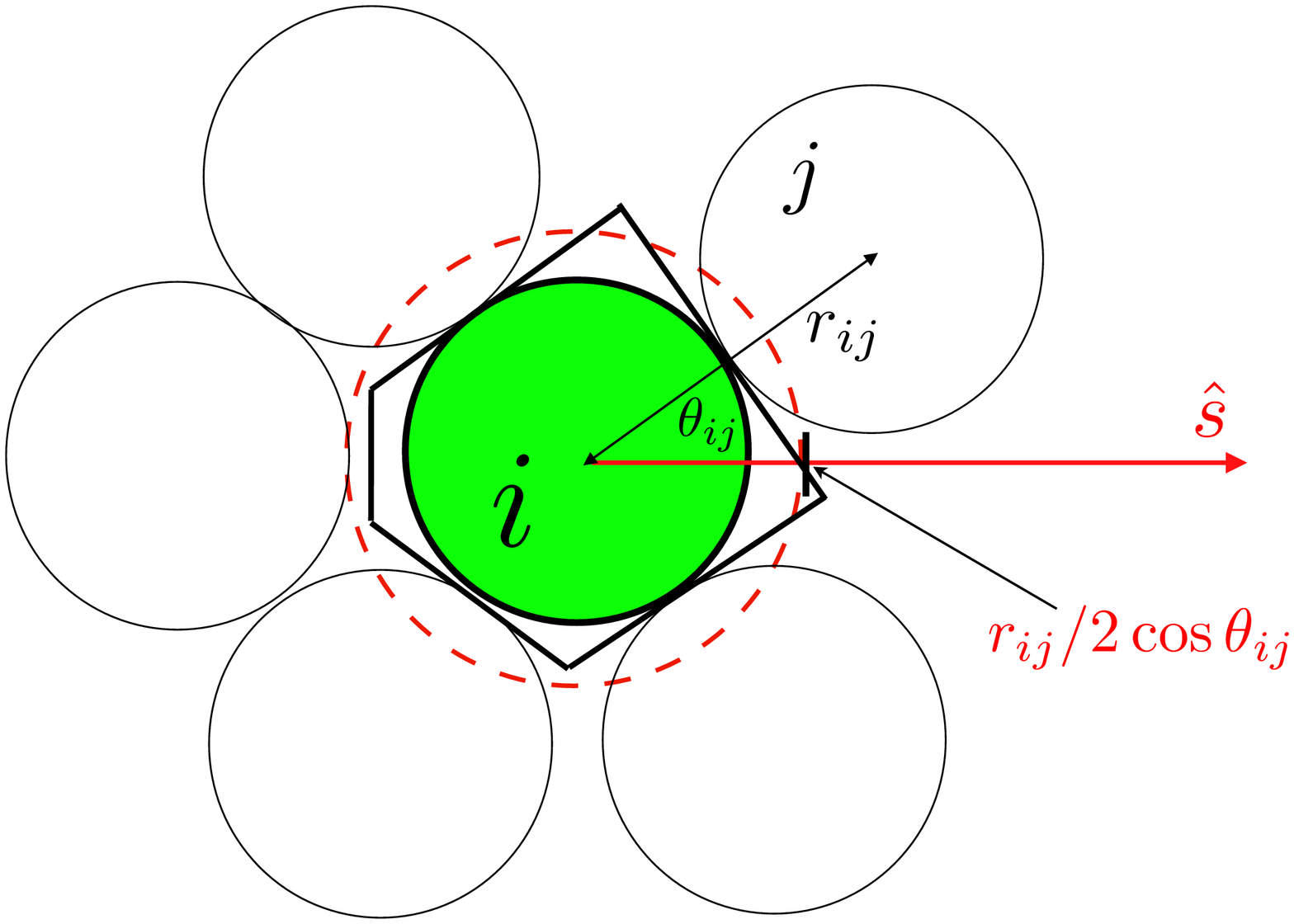}} }} \vspace{.5cm}
  \caption{ Schematics of the Voronoi volume and the orientational
    volume associated with particle $i$.  The boundary of the Voronoi
    cell (shown in 2d for simplicity) corresponds to the irregular
    pentagon in black which defines ${\cal W}_i^{\rm vor}$. The limit
    of the Voronoi cell of particle $i$ in the direction $\s$ is the
    minimum of $r_{ij}/2 \cos \theta_{ij}$ over all the particles in
    the packing, as indicated. This defines the orientational volume
    ${\cal W}_i^s$ which is the volume of the sphere of radius
    $r_{ij}/2 \cos \theta_{ij}$ defined by the dash red circle in the
    figure.  The Voronoi volume is the integration of the
    orientational volume over $\s$ as in Eq.  (\ref{wvor}). Notice
    that $w^s$, the orientational free volume associated with ${\cal
      W}^s_i$ defined in Eq. (\ref{reduced}), ranges from zero (when
    the orientational volume coincides with the volume of the central
    ball, that is when the direction $\s$ coincides with a contact
    point) to in principle $\infty$ for an isolated ball, although the
    maximum $w^s$ in a jammed system is, of course, bounded by the
    free space given by the first or second coordination shell. The
    Voronoi free volume, on the other hand, cannot be zero, by
    definition.}
\label{voronoi-figure}
\end{figure}

To calculate the volume distribution using the established mesoscopic formalism, we start by introducing the different definitions of the volume
associated with each particle necessary to understand the problem.
The starting point is the volume of a Voronoi cell associated to
each particle. The Voronoi tessellation tiles the entire packing
is shown in \cite{jamming1,jamming2} to be a good candidate for
the volume function of jammed matter.  The volume function
replaces the Hamiltonian in thermal systems, and describes the
state of the jammed packings in the ensemble average in the
partition function \cite{sirsam,jamming2}. The Voronoi volume for
each particle ${\cal W}_i^{\rm vor}$ gives rise to the total
volume of the system ${\cal W} = \sum_{i=1}^N {\cal W}_i^{\rm
vor}$, when summed up over all the $N$ particles. In terms of the
relative coordinates of the particles, $\vrij$, we have obtained
in \cite{jamming1,jamming2} for monodisperse particles of radius
$R$ and volume $V_g$ the following formula for the Voronoi volume:

\begin{equation} \label{vor1} {\cal W}_i^{\rm vor} = \frac{1}{3} \oint
  \left(\min_{\s\cdot\drij > 0}(\frac{\rij}{2 \s\cdot\drij })\right)^3
  ds.
\end{equation}
where the integration is performed over the direction $\s$ forming an
angle $\theta_{ij}$ with $\vrij$ as in Fig. \ref{voronoi-figure},
 and $\cos\theta_{ij} = \s\cdot\drij$.  Taking advantage of this
integration we can define an orientational Voronoi volume, ${\cal
  W}_i^s$, for a fixed direction $\s$, satisfying:

\begin{equation} {\cal W}_i^{\rm vor} = \frac{1}{\oint ds} \oint {\cal
    W}_i^s ds = \langle {\cal W}_i^s \rangle_s,
\label{wvor}
\end{equation}
from which we obtain:
\begin{equation}
  {\cal W}_i^s \equiv V_g \left(\frac{1}{2R} \min_{\s\cdot\drij > 0}
    \frac{\rij}{\s\cdot\drij}\right) ^ 3.
\label{voronoi_s}
\end{equation}
${\cal W}^s_i$ defines the orientational Voronoi volume which is
obtained without the integration over $\s$.

The average of the orientational volume over $\s$ for a single
particle, Eq. (\ref{voronoi_s}), is the Voronoi volume, $\langle {\cal
  W}_i^s\rangle_s = {\cal W}_i^{\rm vor}$ and the average of the
orientational volume over many particles for a fixed $\s$ is the same
as the average of the Voronoi volume over the particles: $\langle
{\cal W}_i^s\rangle_i = \langle {\cal W}_i^{\rm vor}\rangle_i$, in the
case of isotropic systems.  This last property is useful since it
allows the use of the orientational volume to define the ensemble
average of the volume fraction without resorting to the use of the
Voronoi volume which contains the average over $\s$ and therefore is
more difficult to treat from a theoretical point of view.  We
therefore promote the use of the orientational volume function ${\cal
  W}_i^s$ as the fundamental quantity to characterize the state of
jammed matter instead of ${\cal W}_i^{\rm vor}$.  It is important to
note that the probability densities of $P({\cal W}^s_i)$ and $P({\cal
  W}^{\rm vor}_i)$ in general differ. For instance, as discussed in
Fig. \ref{voronoi-figure} the orientational free volume can be for
instance zero while the Voronoi free volume cannot. The distribution of Voronoi volume $P({\cal
  W}^{\rm vor}_i)$ can be fitted by a Gamma distribution \cite{Aste2007}, however, $P({\cal W}^s_i)$
has a different form as shown below.

We define the reduced free orientational volume function as
\begin{equation}
w^s \equiv \frac{{\cal W}_i^s - V_g}{V_g},
\label{reduced}
\end{equation}
(we drop the subscript $i$ in $w^s$ for simplicity of notation).

In what follows, we provide a theory for the probability
distribution function of the orientational free volume, $P(w^s)$,
which is less complex than the full Voronoi volume distribution of
Eq. (\ref{vor1}). In \cite{jamming1}, this distribution is
obtained under assumption of uniformity in the packing, making the
theory valid at a mesoscopic level of a few particles diameters.
This approximation can be seen as defining quasiparticles of free
volume $w^s$ capturing the behavior at the mesoscopic distance.
Under this approximation the inverse cumulative distribution $P_>$
is obtained (see Eq. (20) in \cite{jamming2} and Eq. (41) in
\cite{jamming1}) from where the probability density can be
calculated as, $ P(w^s) =\frac{d(1-P_>)}{dw^s}$, then

\begin{equation}
  P(w^s) = \frac{1}{w} \exp\left(-\frac{w^s}{w}\right),
\label{pws}
\end{equation}
where the average value over the particles,
\begin{equation}
  w \equiv \langle w^s\rangle_i = \int w^s P(w^s) dw^s,
\end{equation}
was found to be directly related to the geometrical
coordination number $z$ (Eq. (\ref{eq:w1})) as :
\begin{equation}
  w(z)= \frac{\kappa}{z},
  \label{phi3}
\end{equation}
where $\kappa=2\sqrt{3}$.

We note that
\begin{equation}
\langle w^s\rangle_i =  \frac{\langle{\cal W}_i^s \rangle_i}{V_g}-1
= \frac{\langle{\cal W}_i^{\rm vor} \rangle_i}{V_g}-1.
\end{equation}
Therefore, the orientational volume $w^s$ captures the behavior of the
average volume function and thus can be used as the fundamental
variable to define the microstates of the system instead of the more
complicated Voronoi volume.


The distribution of Eq. (\ref{pws}) is not the distribution that
one would obtain in real packings (generated either experimentally
or numerically) corresponding to the ensemble average of Eq.
(\ref{pws}).  Therefore, further examination is required to derive
the distribution of orientational volumes which can be directly
compared with real packings; their states determined by the
compactivity, $X$.

Under the volume ensemble point of view \cite{sirsam}, the observables
in real packings are ensemble average over the Boltzmann distribution
function.
Using Eq. (\ref{eq:ensAve}), the probability
distribution of volumes in the canonical volume ensemble for a single
quasiparticle of orientational volume $w$ is then:

\begin{equation}
\begin{split}
  P(w^s|X,Z) &= \frac{1}{{\cal Z}_{\rm iso}} \int_{Z}^{6} P(w^s) \exp\left
    [-\frac{w(z)}{X} \right] g(z) dz\\
  &= \frac{1}{{\cal Z}_{\rm iso}} \int_{Z}^{6}
  \frac{1}{w(z)}\exp\left[-\frac{w^s}{w(z)}\right]
  \exp\left [-\frac{w(z)}{X} + \frac{2 \sqrt{3}}{w(z)} \ln h_z\right] dz,
\end{split}
\label{p1}
\end{equation}
where we have used the inverse relation Eq. (\ref{phi3}) and the exponential density of states $g(z)\sim (h_z)^z$.



\begin{figure}
 \centering {
\resizebox{12cm}{!} { \includegraphics{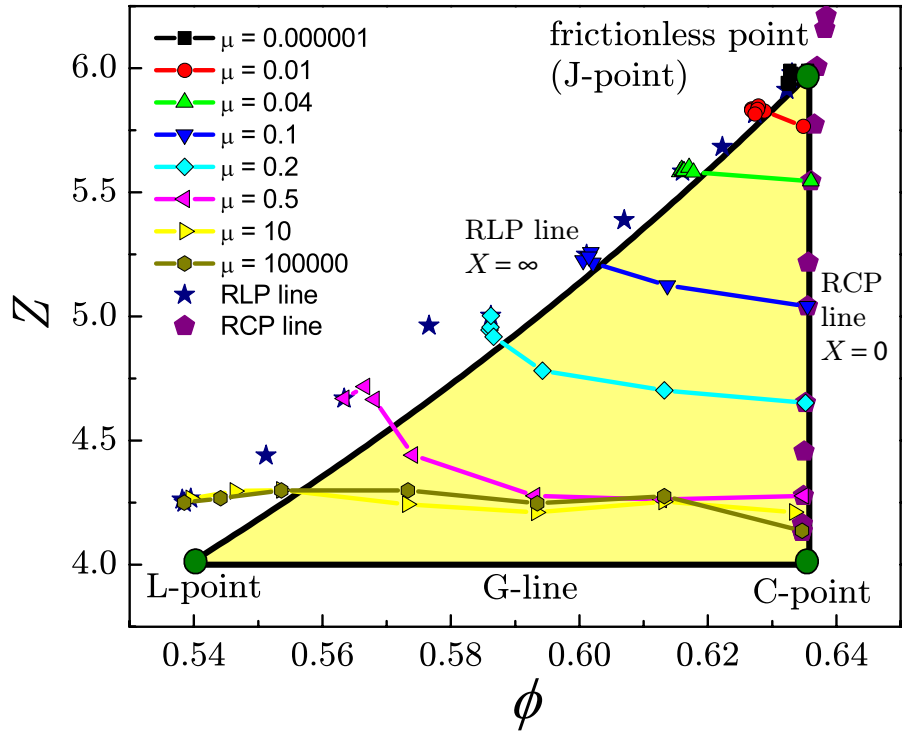}}
}
\caption {Computer generated packings arranged in the phase diagram of
  \cite{jamming2}.  The packings are used to calculate the
  distribution of orientational volumes and coordination number. We
  concentrate our study on packings generated with different friction
  as indicated in the figure following the RLP line from the J point
  to the L point and the RCP line from the J point to the C point.}
\label{phase}
\end{figure}

This equation cannot be solved analytically for a general $(X, Z)$.
However, analytical forms can be obtained in the limiting cases of
$X=0$ (the ground state) and $X\to\infty$: the RCP and RLP lines in
the terminology of \cite{jamming2} respectively (see Fig.~\ref{phase}).
The advantage of studying these distributions is that they can be
checked with simulations or experiments without the use of the
compactivity as a fitting parameter.

From Eq. (\ref{p1}), we find along the RCP line, $P_{\rm RCP}(w^s|Z)
\equiv P(w^s|X=0,Z)$:
\begin{equation}
  P_{\rm RCP}(w^s|Z) = \sqrt{3} \exp\Big(-w^s\sqrt{3}\Big),\,\,\,\,
  Z\in[4,6],
\label{p_rcp}
\end{equation}
and for the RLP line, $P_{\rm RLP}(w^s|Z) \equiv P(w^s|X\to\infty,Z)$:
\begin{equation}
 P_{\rm RLP}(w^s|Z) = \frac{Z}{2 \sqrt{3}} \exp\Big(-\frac{w^s
  Z}{2\sqrt{3}}\Big),\,\,\,\, Z\in[4,6].
\label{p_rlp}
\end{equation}
We note that both limiting distributions coincide at $Z=6$, the
frictionless J-point.

In what follows, we test the above predictions with computer
simulations.  We generate packings at the jamming transition using
the split algorithm explained in \cite{jamming2}. The packings
consist of 10,000 spherical equal-size soft particles interacting
via Hertz normal forces, Mindlin tangential forces and the Coulomb
condition with friction coefficient $\mu$.  The mechanical
coordination number $Z$ versus the volume fraction $\phi$ of the
generated packings are plotted in Fig. \ref{phase} in the
framework of the phase diagram of \cite{jamming2}.  We change
friction from $\mu=0$ to $\mu\to \infty$ to generate the packings
along the RLP line as indicated in the figure with the
corresponding change in the mechanical coordination number from
$Z(0)=6$ to $Z(\infty)=4$. The RCP line is also generated by
changing friction but the volume fraction remains constant, as
seen in the figure, while the mechanical coordination varies from
6 to 4.

\begin{figure}
 (a)  \centering { \resizebox{6cm}{!}{\includegraphics{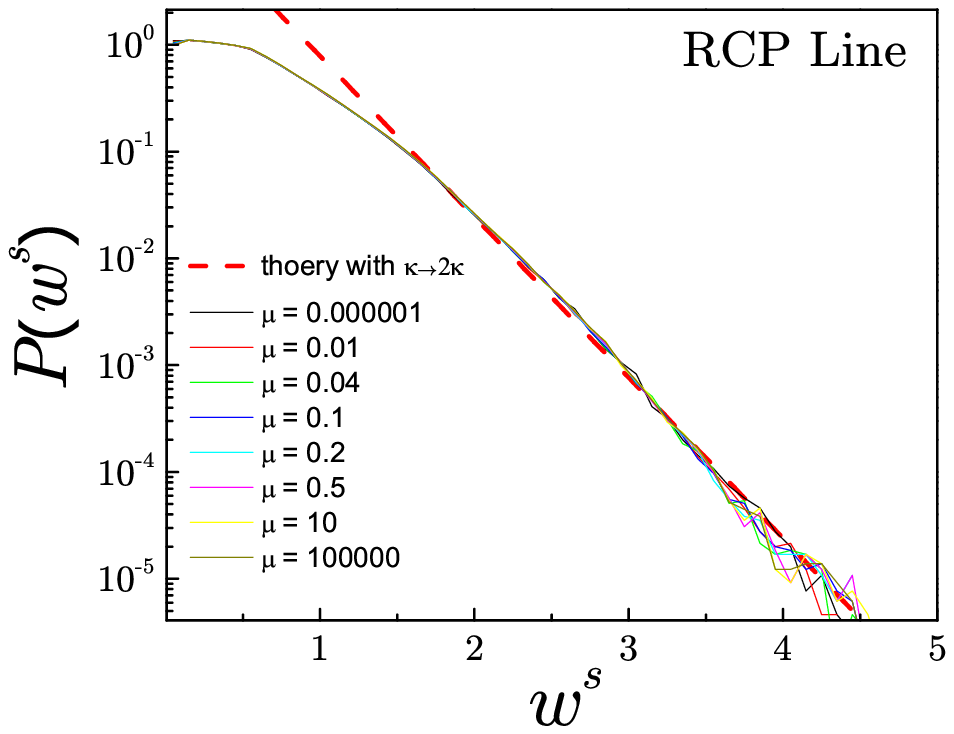}}
}
\centerline{
(b)    \resizebox{6cm}{!}{\includegraphics{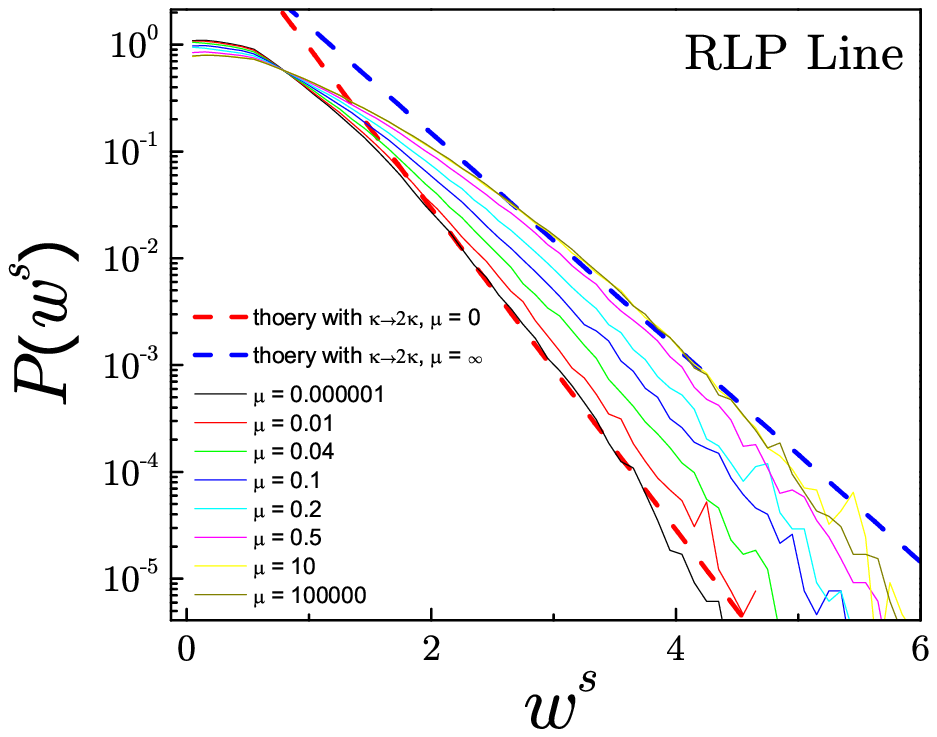}} }
\caption{PDF of $w^s$. Exponential fit in a semilog plot.  (a) We plot
  the results for the distribution of the packings along the RCP line
  with different values of friction, as indicated in the figure (see
  Fig \ref{phase}). The red dashed line is a fit with the theoretical
  prediction $P_{\rm RCP}(w^s)$, Eq. (\ref{p_rcp}), but with the
  inverse characteristic volume or slope of $2 \sqrt{3}$ instead of
  $\sqrt{3}$ as predicted by the theory.  (b) Same as (a) but for the
  packings along the RLP line in Fig. \ref{phase} prepared with
  different $\mu$. The red and blue dashed lines are fits to the
  theoretical prediction $P_{\rm RLP}(w^s)$, Eq. (\ref{p_rlp}), but
  with the inverse characteristic volume or slope replaced from $Z/(2
  \sqrt{3})\to Z/\sqrt{3}$, twice as predicted by the theory like in
  (a). The red line is for $\mu=0$ and has slope $2\sqrt{3}$.  The blue
  line is for $\mu\to\infty$ and has slope $4/\sqrt{3}$.  }
\label{pw}
\end{figure}

We focus on the calculation of the probability distribution
function of $w^s$ for the packings along the RCP-line to test Eq.
(\ref{p_rcp}) and along the RLP-line to test Eq. (\ref{p_rlp}).
Figure \ref{pw} shows the results.  Along the RCP-line, shown in
Fig. \ref{pw}a, we find all distributions are the same,
independent of friction and $Z$, as suggested by Eq.
(\ref{p_rcp}). On the other hand, the distribution along the RLP
line, shown in Fig. \ref{pw}b, depends on friction and therefore
on $Z(\mu)$ as suggested by theory. The exponential dependence
seems to be captured upon a first inspection of the data done in a
semi-log graph of Figs. \ref{pw}a and b (arguably better for the
RCP case), at least for intermediate values and the tail of the
distribution. In the case of the RLP line (Fig. \ref{pw}b), the
slope of the semi-log plot of the exponential fit has the same
dependence on $Z$ as predicted by theory (that is the slope in the
semi-log plot increases linearly with $Z$) but with twice the
constant value as predicted by Eq. (\ref{p_rlp}). We find that the
exponential fit leads to a tail with characteristic volume
$=Z/\sqrt{3}$, twice the value predicted by theory
$Z/(2\sqrt{3})$, Eq. (\ref{p_rlp}). However, the linear trend with
$Z$ is observed in the data.

On the other hand, the average of the distributions agrees very
well with simulations (see below). But when we fix the mean
according to theory the exponential tail is inaccurate.  Theory
either provides the exponential fit with the incorrect average or
provides the correct average with the deviations from the
exponential fit.

The same situation is observed in Fig. \ref{pw}a for the RCP line. We
force the fitting to be exponential, and then the average value has to
be modified from $\sqrt{3}\to 2\sqrt{3}$.  The reason for this
discrepancy is that the theory does not capture the distribution in
the full range of volumes.
Figure \ref{pw} clearly show a plateau at small values of $w^s$
deviating from the exponential behavior predicted by the theory.

Furthermore, a more strict scrutiny of the data seems to indicate
that the exponential fit may not be sufficiently accurate as shown
in Figs. \ref{pw}a and b. While tempting to conclude that the
exponential is a good fit to the data (for instance, the fitting
in Fig. \ref{pw}a looks convincing), further scrutiny reveals
important deviations in the tail. To visualize the deviations, one
should take the plot and look at it, not frontally, but from the
side. It is evident that there is a slight curvature in the
distributions deviating from the linearity in the semi-log plots.
Indeed, the distributions decay slightly faster than the pure
exponential decay predicted by theory.  The largest evidence of
this deviation is perhaps in the tail of the $\mu\to \infty$ data
in the RLP, Fig. \ref{pw}b.

\begin{figure}
\centering {\vbox{
(a)     \resizebox{6cm}{!}{\includegraphics{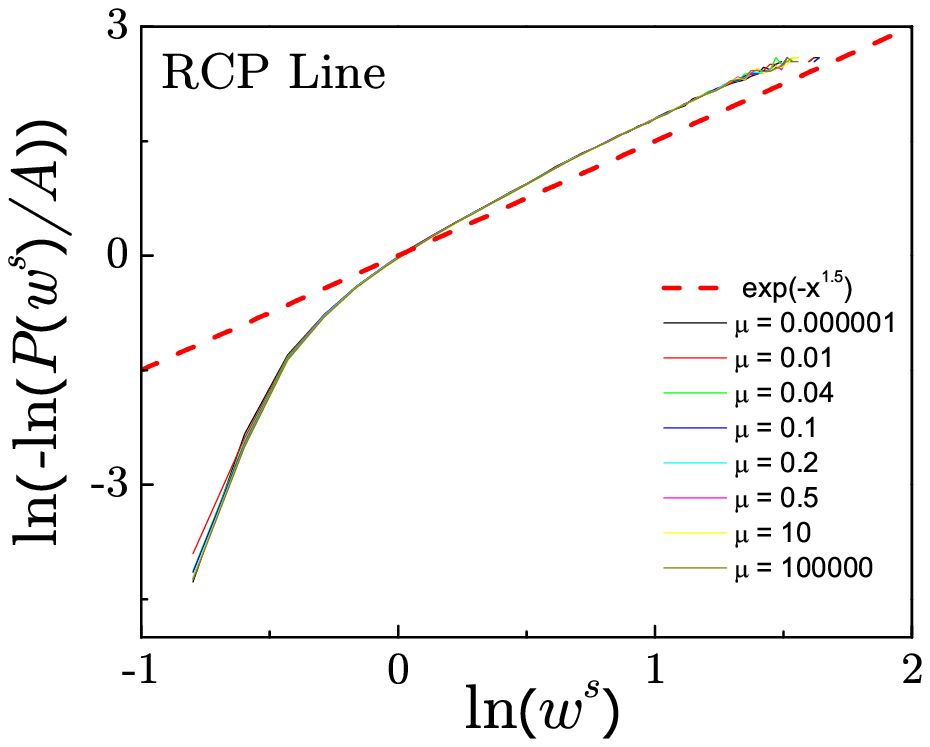}}
}}
\centering {\vbox{
(b)     \resizebox{6cm}{!}{\includegraphics{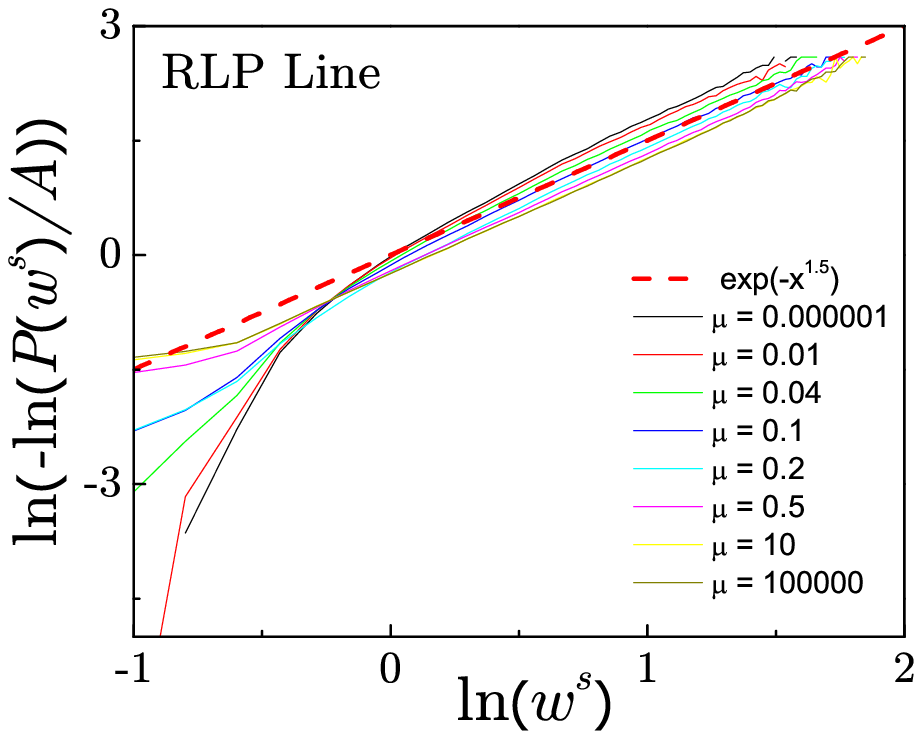}} }}
\caption{PDF of $w^s$. Compressed exponential fit in a double log
  plot.  The results are the same as in Fig. \ref{pw} but now reploted
  in a double log plot to obtain the compressed exponential fitting
  exponent $\beta$ which is extracted from the sloe of such a plot, as
  explained in the text. (a) PDF for the packings along RLP-line. (b)
  PDF for the packings along the RLP-line.}
\label{lnpw}
\end{figure}

A double log analysis of the data shown in Figs. \ref{lnpw}a and
\ref{lnpw}b reveals that a compressed exponential behaviour might
better capture the tail of the distributions above the average
value:

\begin{equation}
  P(w^s|Z) \sim A \exp\Big[-\Big(\frac{w^s}{w_c}\Big)^{\beta_w}\Big], \,\,\,\, w^s>w,
\label{compr}
\end{equation}
where $\beta_w\approx 1.5$ is the compressed exponential exponent
($\beta_w=1$ would be a pure exponential) valid for all the RCP
and RLP packings according to Fig. \ref{lnpw}a and \ref{lnpw}b,
$w_c$ is a characteristic volume independent of $Z$ in the RCP
packings and depends on $Z$ for the RLP packings, and $A$ is a
constant.

We want to stress the difficulties associated with a fit to a
compressed (or stretched) exponential function like Eq. (\ref{compr}).
A double log plot gives:
\begin{equation}
  \ln\Big(-\ln\big(P(w^s|Z)/A\big)\Big) = \beta_w \ln(w^s) - \beta_w \ln(w_c),
\label{comp2r}
\end{equation}
providing a linear fit with slope $\beta_w$. Beyond the inherent
subtleties associated to taking a double log of a function in a such a
short range, a further complication arises because such a linear fit
depends on the value of the constant $A$. Since Eq.  (\ref{compr})
covers only the tail of the distribution, $A$ cannot be obtained from
normalization, remaining as a fitting parameter.  The result is a
fitting of $\beta_w$ dependent slightly on $A$, providing an extra
level of difficulty.

Although the compressed exponential seems to fit the data better
than the pure exponential, before more theory or
numerical/experimental evidence in the limit $N\to\infty$ become
available, we are inclined to conclude that the problem is not
closed.  We note that a similar dichotomy between exponential and
compressed/stretched exponential behavior has plagued the study of
the distribution of forces in jammed matter since the first
studies on the subject \cite{Tighe2010}. Given the inherent difficulties in any
numerical estimation, the dispute will have to eventually be
settled when more exact theories become available.

Beyond the distributions of volumes, the theory reproduces very
well the average value of the volumes, Eq. (\ref{phi3}). Based on
the properties of the averages expressed above, there is no need
to calculate the full Voronoi volume to obtain the average volume
fraction, since the average of the orientational $w^s$ suffices.
For instance in the frictionless packing we find $\langle
w^s\rangle = 0.561$, which gives a volume fraction $\phi =
1/(1+\langle w^s\rangle) = 0.641$, in agreement with the direct
measurement of the volume fraction of the packing, 0.64.

\begin{figure}
  \centering { \vbox { (a)
      \resizebox{6cm}{!}{\includegraphics{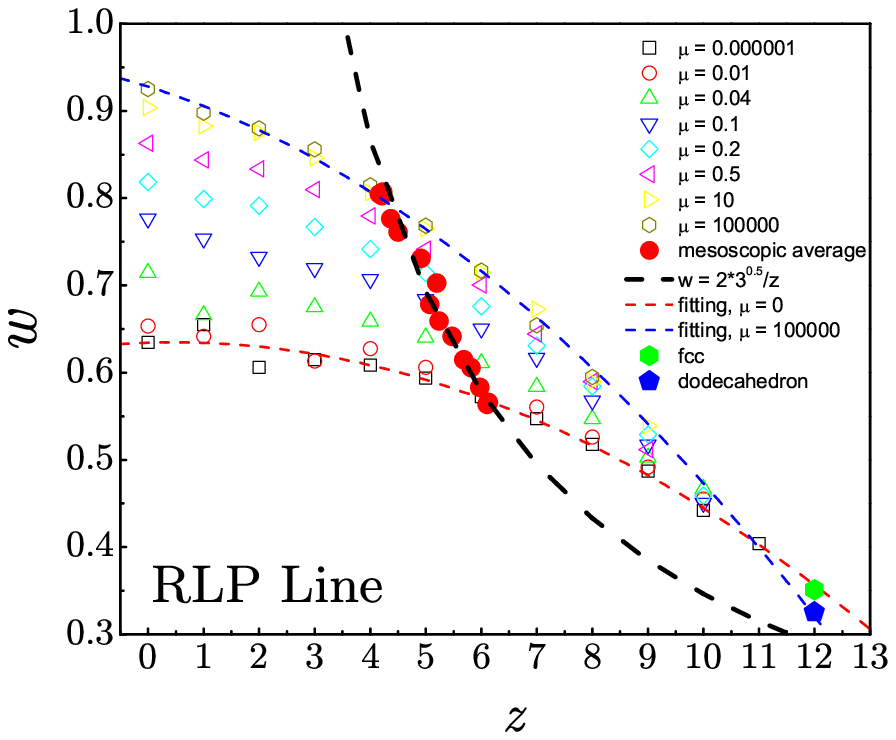}}
(b)\resizebox{6cm}{!}{\includegraphics{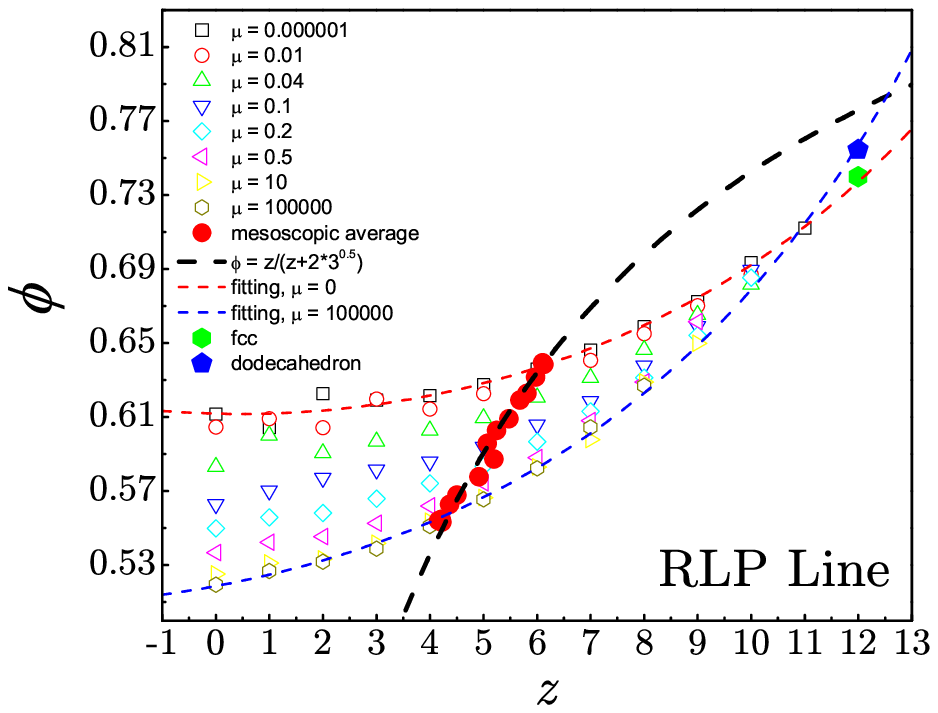}} } }
\caption{Relation between local volume per particle and its
  coordination number for (a) the volume function and (b) the volume
  fraction. We use the packings along the RLP line with the friction
  as indicated and we calculate the volume function binning the data by
  the coordination number of the particles. The red dots correspond to
  the average over all the particles in the packing of the volume
  function which fits the theory very well. The red and blue dashed
  lines are logarithmic-like fittings to the data and we add the value
  at $z=12$ for the FCC and the dodecahedron. The data from the
  packings goes only up to $z=11$, indicating the absence of ordered
  structures.}
\label{wz}
\end{figure}

A full comparison between theory and simulations is given in Fig.
\ref{wz}, where we study the dependence between average volume and
coordination number.  For each packing along the RLP line we
calculate the average orientational volume focusing on the
particles with a given $z$. We also calculate the average over all
the particles for a given packing, plotted as the red dots in Fig
\ref{wz}.  In practice, we do not measure the geometrical
coordination number $z$ but the mechanical coordination number
$Z$.  However, we know that for the packings along the RLP line
$z\approx Z$ \cite{jamming2} (this is because $h_z\to 0$).
Furthermore, the RLP line corresponds to $X\to \infty$ and
therefore the prediction of the average volume function, Eq.
(\ref{phi3}) can be tested directly with these fully random
numerical packings, extending this result, valid for a
quasiparticle, to the entire packing. Thus, the packings along the
RLP line reveal the approximate behaviour of quasiparticles of
fixed coordination $z$. We notice that there could be still some
subtleties when comparing Eq. (\ref{phi3}), valid for
quasiparticles, to the results in packings. Using $\phi^{-1}=
w+1$, we plot the volume fraction in Fig. \ref{wz}b.

Figures \ref{wz}a shows that the mean of the distribution of
volumes, $w$, is well captured by the theory of Eq. (\ref{phi3})
(see the black dashed line in comparison with the red dots in Fig.
\ref{wz}).  This is why the theory provides very good fittings to
the values of RCP and RLP in \cite{jamming2}. The agreement can be
seen as well in the volume fraction in Fig. \ref{wz}b, and exists
despite the fact that the full distribution presents the
deviations discussed above.

Figure \ref{wz} presents further interesting results.  For a given
packing along the RLP-line specified by a fixed friction, there
are a variety of particles with varying coordination $z$,
following a well-defined functional relation between the volume
occupied by the particle and its coordination number (see for
instance the red and blue dashed lines in Fig. \ref{wz}a
corresponding to fittings for the cases $\mu=0$ and $\mu \to
\infty$, respectively). While this plot does not tell us how many
particles there are for a given coordination number (see next
section) we see that for each packing, there exists a variety of
local volume functions with $z$ ranging from $z=0$ (since there
are some rattlers) up to $z=11$ (but not 12, interestingly, see
below).

The assumptions for the limits of integration in the partition
function in \cite{jamming2} or the ensemble average Eq.
(\ref{eq:part2}), $Z\le z\le6$, seem to be violated here. However, we
have to remember that the theory is mesoscopic and further
coarse-graining is needed for these bounds to be more accurate.
Regardless, even though the range in $z$ extends further than the
bounds, Fig. \ref{wz} corresponds to the average for a fixed $z$
but does not tell how many states there are for every $z$.  When
these details are properly taken into account, the bounds are
approximately satisfied, although fluctuations persist, bringing
us to the next Section of this paper.

Focusing around the $z=12$ point in the figures, we observe that an
extrapolation of the fitting to the curve of frictionless packing
seems to converge to the green dot in Figs.  \ref{wz} at $z=12$ which
correspond to the free volume function of FCC, $w_{\rm FCC}= 0.35135$
(Fig. \ref{wz}a) and the FCC volume fraction $\phi_{\rm FCC}=
\pi/\sqrt{18} \approx 0.7402$ (Fig. \ref{wz}b).  An extrapolation of
the fitting to the infinite friction data seems to pass through the
volume fraction of the dodecahedron as indicated by the blue point in
Fig. \ref{wz}, which has also $z=12$ but slightly larger volume
fraction that FCC (the dodecahedron can't tile the space without
leaving holes, so the best global packing is still the FCC).  We
observe that there are no particles in the packings with $z=12$.
Indeed the green and blue points at $z=12$ in Figs.  \ref{wz}a and b
were added by hand and the real curves stop shortly at $z=11$. The
absence of $z=12$ states indicate the randomized state of the systems.

More importantly, we see that if we extend the theoretical result of
Eq. (\ref{phi3}) to the ordered region, examining $z$ from $z=6$ to
$z= 12$, the theory does not fit the FCC value. Instead, we obtain
$w(z=12)= 2\sqrt{3}/12 = 0.2886$, below the FCC or dodecahedron
value. In principle, this result is expected since the theory assumes
random isotropic states while the FCC is an ordered anisotropic
packing. However, the absence of a good fitting of the disordered
branch through the FCC, together with the fact that the theory fits so
well the disordered states, raises the interesting question of the
existence of a phase transition between the RCP limit at $z=6$ and the
FCC at $z=12$. Since packings cannot equilibrate above $z=6$ without
the formation of crystalline regions, we expect an ordered branch from
the FCC point towards the RCP point.  It seems plausible that there
could be a discontinuity from the disordered branch to the ordered
branch, the existence of which could determine whether there exist a
disorder/order phase transition characterizing the RCP.
This scenario has been confirmed by analysis of numerical packings in a recent study \cite{Jin2010}, which showed that RCP can be interpreted as a ``freezing point" in a first-order phase transition between ordered and disordered phases.

To summarize this part of the study, while theory predicts an
exponential behavior and approximates well some features of the
distribution such as the average value, simulations indicate that
a compressed exponential fitting could be also possible. We
therefore require more refined theories to account for the full
behavior of the volume distribution. The present approach suggests
that study of the ensemble of quasiparticles of fixed coordination
number could provide clues to the behavior of the entire system
through the Edwards ensemble approach.  While new theoretical
concepts are required, current attempts indicate that it might be
possible to develop a theory of the volume distribution that is
exact to, at least, a given coordination shell of particles.  It
appears that $P(w^s)$ for a fixed $z$-ensemble might be solved
exactly by a brute force approach, since the range of a Voronoi
cell is finite. Although, it may contain a large number of
variables, the computer should handle such a computation. Such an
analysis parallels the Hales proofs of the Kepler's conjecture
\cite{hales}.


\section{PDF of the coordination number}

Next, we analyze the distribution of coordination number by
generalizing the theory of \cite{jamming2} to include fluctuations in
$z$.  While we have assumed \cite{jamming2} that every single
quasiparticle satisfies the specified bounds: $Z\le z \le 6$, below we
relax this constraint to extend the bounds to the geometrical
coordination of the entire system by considering:

\begin{equation}Nz_{\rm min} \le \sum_{i=1}^{N} z_i \le Nz_{\rm max},
\end{equation}
where in the following we set $z_{\rm min} = Z$ and $z_{\rm max} = 6$.
This new condition implies that it is not possible to consider the single
quasiparticle partition function, Eq. (\ref{eq:part2}),
and that the full $N$-particle partition function has
to be considered:
\begin{equation}
\begin{split}
& \mathcal{Z}=\int\ldots\int_{Nz_{\rm min} \le
    \sum_{i=1}^{N} z_i \le Nz_{\rm max}}
\prod_{i=1}^{N} e^{-w_i(z_i)}g_i(z_i)dz_i \\
&=\int\ldots\int_{Nz_{\rm min} \le
    \sum_{i=1}^{N} z_i \le Nz_{\rm max}}
\prod_{i=1}^{N} e^{-(z_i/z^* + \beta\kappa/z_i)}dz_i.
\end{split}
\end{equation}
We have $z^*=-1/\ln h_z$, the inverse compactivity $\beta=1/X$ and
the coupling constant
\begin{equation}
B=\frac{\beta \kappa}{z^*}.
\label{b}
\end{equation}
Then, the ensemble average of the probability
distribution function of the geometrical coordination number, $P(z)$,
is
\begin{equation}
\begin{split}
  P(z)  & \equiv \left< \frac{1}{N} \sum_{i = 1}^{N} \delta(z-z_i)
  \right>  \\
   &= \frac{1}{\mathcal{Z}}\int\ldots\int_{(Nz_{\rm min}-z) \le
    \sum_{i=1}^{N-1} z_i \le (Nz_{\rm max} - z)}
   \prod_{i=1}^{N-1} e^{-(z_i/z^* + \beta\kappa/z_i)}dz_i,
\end{split}
\end{equation}

 We obtain:

\begin{equation}
P(z)  =   \frac{\mathrm{Erf}(\frac{\sqrt{N-1}(z_{\rm max}'- \mu)}{\sqrt{2}\sigma})
    + \mathrm{Erf}(\frac{\sqrt{N-1}( \mu-z_{\rm min}')}{\sqrt{2}\sigma})}
  {\mathrm{Erf}(\frac{\sqrt{N}(z_{\rm max}- \mu)}{\sqrt{2}\sigma}) +
    \mathrm{Erf}(\frac{\sqrt{N}( \mu-z_{\rm min})}{\sqrt{2}\sigma})}
   e^{-(z/z^* + \beta\kappa/z)},
\label{erf}
\end{equation}
with the following constants:

\begin{equation}
\begin{split}
z_{\rm max}' \equiv (Nz_{\rm max}-z)/(N-1) \approx z_{\rm max} +
(z_{\rm max} - z)/N ,\\
z_{\rm min}' \equiv (Nz_{\rm min}-z)/(N-1) \approx
z_{\rm min} + (z_{\rm min} - z)/N,
\end{split}
\end{equation}
and $\mathrm{Erf}(x)$ the Gauss error function.

The constants $\mu$ and $\sigma$ in Eq. (\ref{erf}) are
significant because they represent the mean and standard deviation of
the a Gaussian expansion in a saddle-point approximation of the
inverse Fourier transform of the partition function allowing the
calculation of the free-volume density.  They are:
\begin{equation}
\mu = z^*B^{1/2}\frac{K_1(2 B^{1/2})}{K_0(2 B^{1/2})},
\end{equation}
which is the same as the ensemble average of the coordination number, and
\begin{equation}
\sigma^2 \equiv {z^*} ^{2} B\left(\frac{K_2(2 B^{1/2})}{K_0(2
B^{1/2})} - \frac{K_1(2 B^{1/2})^2}{K_0(2 B^{1/2})^2}\right),
\end{equation}
where $K_n(a)$ is the modified Bessel function of the second kind:
\begin{equation}
\int_0^{\infty} x^n e^{-\frac{a}{2}(x+1/x)} dx = 2 K_n(a).
\end{equation}




Next, we consider the approximations of Eq.~(\ref{erf}) for two cases: When $z_{\rm max}-\mu <\mu - z_{\rm min}$, then
$\mu > (z_{\rm max}+z_{\rm min})/2$, and we obtain:
\begin{equation}
P(z) \sim
\exp\left[-\frac{1}{z^*}\left(z\left(2-\bar{w}/w_{\rm min}\right) +
\frac{1}{z}\frac{\kappa^2}{\bar{w}^2}\right)\right],
\label{pz1}
\end{equation}
or otherwise, we obtain:
\begin{equation}
P(z) \sim
\exp\left[-\frac{1}{z^*}\left(z\left(2-\bar{w}/w_{\rm max}\right) +
\frac{1}{z}\frac{\kappa^2}{\bar{w}^2}\right)\right],
\label{pz2}
\end{equation}
where $\bar{w}$ is the ensemble average of the volume function (which depends on $\beta$, or compactivity $X$), $w_{\rm min} = \kappa /z_{\rm max}$ and $w_{\rm max} = \kappa /z_{\rm min}$.

In the following we consider the distribution functions at two special points on the phase diagram and compare them to the numerical simulations. At the frictionless J-point and the infinitely frictional L-point (see
Fig. \ref{phase}), the
distributions reduce to simple forms. For the RCP J-point, $X = 0$ and the system has the minimum average volume and maximum average coordination number, therefore, $\bar{w} \sim w_{\rm min}$ and $u \sim z_{\rm max}$. From Eq.~(\ref{pz1}) We find:
\begin{equation}
  P_{\rm RCP} (z) \sim
  \exp\left[-\frac{1}{z^*} \left(z + \frac{z_{\rm max}^2}{z}\right)\right].
\label{eq:pzJ}
\end{equation}
For the RLP L-point, $X \rightarrow \infty$ and the system has the maximum average volume and minimum average coordination number, therefore, $\bar{w} \sim w_{\rm max}$ and $u \sim z_{\rm min}$. From Eq.~(\ref{pz2}), we find:
\begin{equation}
  P_{\rm RLP} (z) \sim
  \exp\left[-\frac{1}{z^*} \left(z + \frac{z_{\rm min}^2}{z}\right)\right].
\label{eq:pzL}
\end{equation}

\begin{figure}
  \centering { \vbox {
      (a)\resizebox{7cm}{!}{\includegraphics{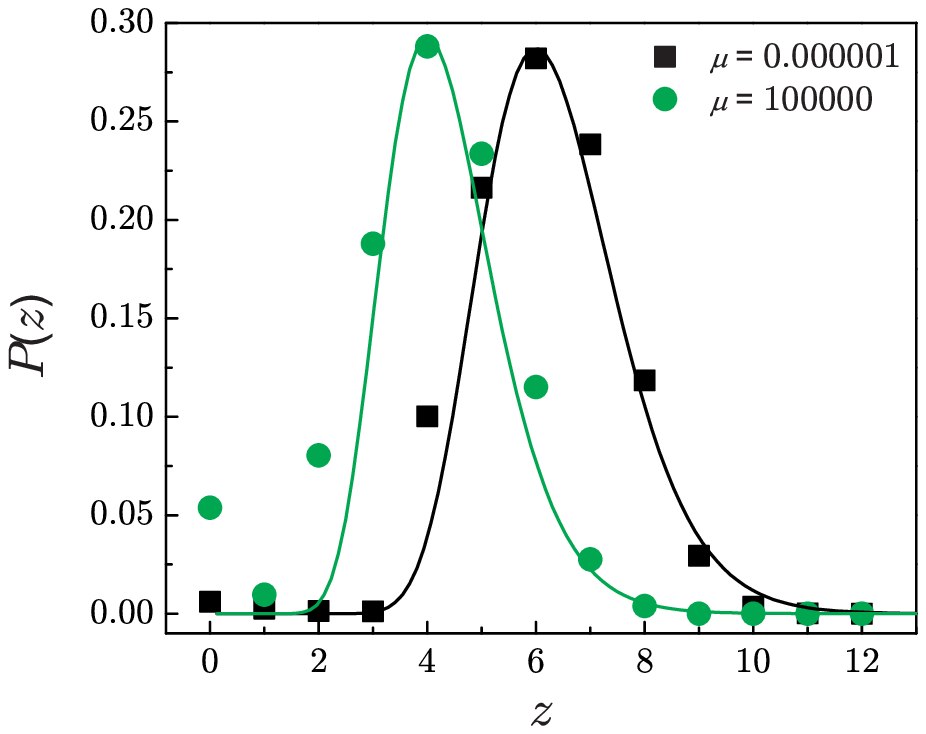}}
      (b)\resizebox{7cm}{!}{\includegraphics{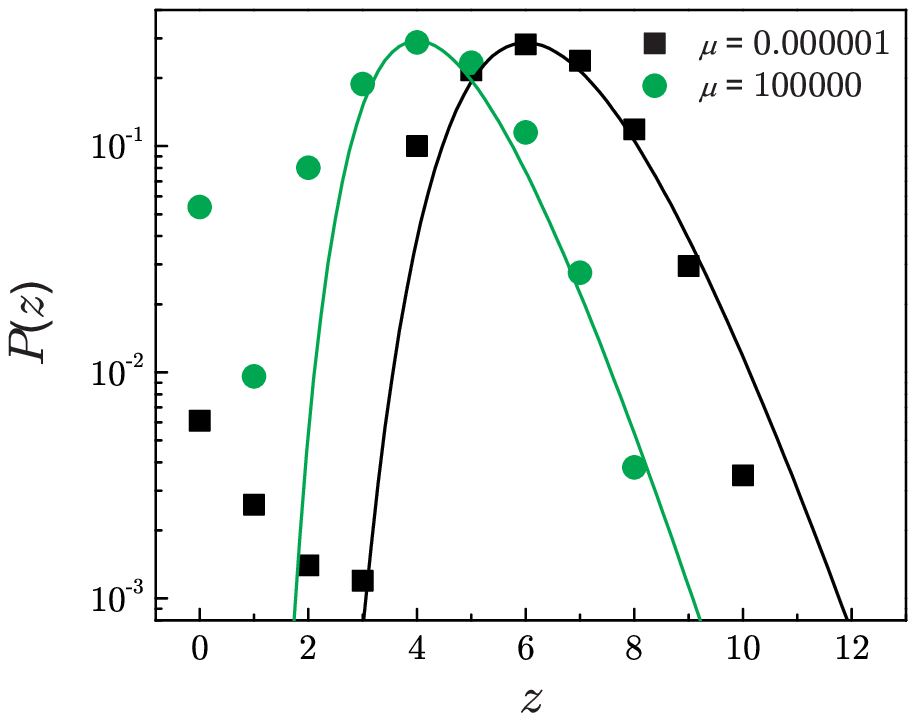}}
}
}
\caption{$P(z)$: comparison between theory and simulations in (a) a
  lin lin plot and (b) a semi-log plot to appreciate better the tail
  of the distribution. We plot the systems at the J-point ($\mu=0$)
  and L-point ($\mu=\infty)$ in Fig. \ref{phase}. we use a fitting parameter $z^* = 0.5$ in Eqs. (\ref{eq:pzJ}) and (\ref{eq:pzL}).}
\label{pz}
\end{figure}

We test these forms with the computer generated packings at the
J-point and L-point. The constant $z^*$ is treated as a fitting parameter since it determines the density of states and is difficult to know a priori. Figure~\ref{pz} shows the result.  The lin-lin
plot of Fig. \ref{pz}a shows that the distribution near the average
value is well approximated by the theory for both points. To
investigate the tails of the distributions, Fig. \ref{pz}b plots a
semi-log curve.  While some deviations are observed, the fit is still
reasonable except for the larger coordination number of J-point and smaller coordination number of L-point, which are very
rare, about $10^{-2}$ less probable than the most probable value.


\section{Conclusions}

We have presented the predictions of the mesoscopic theory
presented in \cite{jamming2} concerning the probability
distribution of the orientational volumes in jammed matter.  The
theory captures very well the average volume and indeed gives rise
to good predictions of the RCP and RLP volume fraction as shown in
\cite{jamming2}.  However, when comparing the full distribution we
find important deviations. For instance, computer simulations are
able to detect slight deviations from the pure exponential
behaviour predicted by theory. This deviation could be better
approximated by a compressed exponential behavior, although a more
conclusive fitting necessitates a more precise theory or
simulations in the large scale limit. The plateau observed in the
volume distribution is not captured by the theory either.  There is evidence that this plateau might originate from the spatial correlations between the first and second layer particles, which indicates that further progress could come from a
systematic analysis of higher level coarse-graining. For example, by explicitly treating the second-layer neighbors, it is possible to improve the distribution functions predicted by the theory. This approach is particularly appropriate for two-dimensional systems, since the free variables are significantly reduced in the 2d case.

On the other hand, the behavior of the probability distribution of
coordination number is captured well by the theory. Here, the
mesoscopic theory of \cite{jamming2}, considering restricted
bounds for each quasiparticle, is extended to allow for
coordination number fluctuations, by imposing the bounds to the
entire system. Such a problem can be solved under several
approximations and predicts a mixed exponential forms that are
well reproduced by the simulations at the J and the L-point.

Why does the theory capture the contact distribution better than the
volume distribution? The distribution of contacts is an ensemble
average based on the states characterized by $w(z)$.  Since this
function is quite accurate, the contact distribution follows. On the
other hand, the volume distribution is based on the uniform
approximations done on \cite{jamming1}, and therefore is not quite
exact. The full distribution of volumes is where more theoretical
developments are required.

\section*{Acknowledgements}
This work is supported by NSF-CMMT, and DOE
Geosciences Division. We thank C. Briscoe and L. K. Gallos for a critical reading of
this manuscript.


\clearpage

\end{document}